\documentstyle[12pt,epsfig]{article}
\newfont{\sfb}{cmssbx10 scaled 1400}
\newfont{\bigsf}{cmssbx10 scaled 1600}

\begin{document}
\begin{flushright}
BIHEP--th--01--01\\
KOBE--FHD--01--02\\
FUT--01--01\\
August~~~2001
\end{flushright}
\begin{center}
{\baselineskip 22pt
{\LARGE\bf Extraction of Polarized Gluon Distributions from
Large--$p_T$ Light Hadron Pair Production}\\
}
\vspace{2.0em}
{
\renewcommand{\thefootnote}{\fnsymbol{footnote}}
Yu-Bing Dong\footnote[2]{E--mail: dongyb@mail.ihep.ac.cn}\\
\vspace{0.8em}
{\it Institute of High Energy Physics, Academia Sinica,}\\
{\it Beijing 100039, P. R. China}\\

\vspace{1em}

T. Morii\footnote[9]{E--mail: morii@kobe-u.ac.jp}\\
\vspace{0.8em}
{\it Faculty of Human Development and}\\
{\it Graduate School of Science and Technology,}\\
{\it Kobe University, Nada, Kobe 657--8501, Japan}\\
\vspace{1em}
and\\
\vspace{1em}
T. Yamanishi\footnote[3]{E--mail: yamanisi@ccmails.fukui-ut.ac.jp}\\
\vspace{0.8em}
{\it Department of Management Science,}\\
{\it Fukui University of Technology, Gakuen, Fukui 910--8505, Japan}\\
\vspace{3.5em}
}
{\bf Abstract}
\end{center}

\baselineskip=22pt

We propose a new formula for extracting the polarized gluon distribution
from the large--$p_T$ light hadron pair production
for semi--inclusive processes in polarized deep inelastic scattering.
In general, a large--$p_T$ hadron pair is produced
via photon--gluon fusion(PGF) and QCD Compton at the lowest order of QCD.
The PGF gives us a direct information on $\Delta g$ in the nucleon,
while QCD Compton becomes background to the signal process for extracting
$\Delta g$.
We show that the contribution from this background, i.e. the QCD Compton
process, can be removed by using
symmetry relation among fragmentation functions and taking an appropriate
combination of various light hadron pair production processes, and
thus the double spin asymmetry can be described in terms of $\Delta g/g$
alone.

\vspace{1.0em}
\noindent
PACS number(s): 13.88.+e, 13.85.Ni, 14.65.Bt

\vfill\eject

\baselineskip 22pt
\noindent

\section{Introduction}

The nucleon is not an elementary but
compound particle.
Accordingly, its spin is carried by its constituents and
is described by a sum rule,
\begin{equation}
\frac{1}{2}=\frac{1}{2}\Delta\Sigma + \Delta g + L_z~,
\label{eqn:1}
\end{equation}
where $\frac{1}{2}$ on the left hand side denotes the
spin of a longitudinally polarized nucleon, while
$\Delta\Sigma$, $\Delta g$ and $L_z$ on the right hand side
represent the amount of the
spin carried by the quarks, gluons and their orbital angular momenta,
respectively.
In the static quark model with SU(6) symmetry,
the nucleon's spin is totally carried by quarks alone and thus,
each term on the right hand
side of eq.(\ref{eqn:1}) becomes $\Delta\Sigma=1$, $\Delta g=0$
and $L_z=0$.
Moreover, in the naive quark--parton model, the value of $\Delta\Sigma$
is predicted to be $\sim 0.7$, which means that the quarks almost
carry the nucleon's spin\cite{Ellis74}.
However, triggered by the first measurement by the EMC in 1988
of the longitudinally polarized structure function of the proton
$g_1^p$ for wide kinematic range of Bjorken $x$ and $Q^2$,
the situation has drastically changed, that is, the
quarks carry the proton spin little and the strange quark
is negatively polarized to the proton spin\cite{EMC88}.
These results are in significant contradiction to
the traditional understandings based on the static quark model and/or
the quark--parton model.
Since then, a great deal of efforts have been made experimentally
and theoretically for disclosing the origin of the nucleon
spin\cite{Review}.
So far, based on the next--to--leading order QCD
analyses on the $x$ and $Q^2$ dependence of
the longitudinally polarized structure function $g_1$,
we have got a number of excellent parametrization models of the
polarized parton distribution functions (pol--PDFs)
from fitting to experimetal data, with high precision, of
polarized deep inelastic scattering (pol--DIS)
for various targets\cite{GRSV96}--\cite{Goto}.
All of these parametrization models
tell us that $\Delta\Sigma$ is around 0.3 or smaller,
that is, the contribution of
the quark spin to the nucleon spin is rather small.
The remainder should be compensated by $\Delta g$ and $L_z$.
To understand the physical ground of these results is the main subject
of the so-called proton spin puzzle, which is still challenging problem
to be solved.
To solve this problem, it is very important
to precisely know how the gluon polarizes in the nucleon.
It is well-known that the next--to--leading order QCD analyses on $g_1$
bring about information on the first moment of the polarized
gluon $\Delta g$\cite{Mertig}.
However, there are large uncertainties in $\Delta g$ extracted
from $g_1$ alone.
Knowledge of $\Delta g$ is still limited because it is
very difficult to directly obtain such information
from existing data.

So far, a number of interesting proposals such as
direct prompt photon production in polarized proton--polarized proton
collisions\cite{Craigie}, open charm\cite{Watson} and
$J/\psi$\cite{Morii} productions
in polarized lepton scattering off polarized nucleon targets,
were presented for studying longitudinally
polarized gluon distributions.
Recently, HERMES group at DESY reported the first
measurement of the polarized gluon distribution from
di--jet analysis of semi--inclusive processes in
pol--DIS, though only one data point was given as a function
of Bjorken $x$\cite{HERMES}.
In general, a large--$p_T$ hadron pair is produced via
photon--gluon fusion (PGF) and QCD Compton at the lowest order
of QCD (Fig. 1).
The PGF gives us a direct information on the
polarized gluon distribution in the nucleon, while QCD Compton
becomes background to the signal process for extracting the polarized
gluon distribution\cite{Bravar,Roeck}.
When we consider only the case of heavy hadron pair productions,
we could safely neglect the contribution of QCD Compton
processes\cite{YDM} because the contents of heavy
quarks in the proton are extremely small and furthermore the probability
of fragmenting of light quarks to heavy hadrons is also small.
However, in this case the cross section itself
is rahter small at the energy in the running(HERMES) or
forthcoming(COMPASS) experiments and thus, we could not have enough
data for doing detailed analysis.  Furthermore,
if we consider only small $x$ region where the PGF dominates over
the QCD Compton, we could also extract the polarized gluon
distribution without difficulty\cite{Review}.
However, here we are interested in the polarized gluon distribution
for wide reigon of $x$ to calculate the first moment
of the polarized gluon.
For light hadron pair productions in wide $x$ region,
QCD Compton contribution is not necessarily small and hence,
it is rather difficult to unambiguously extract the behavior
of the polarized gluon from those processes.

In this work, getting over these obstacles, we propose a new formula
for clearly extracting the polarized gluon
distribution from the light hadron pair production of pol--DIS by
removing the QCD Compton component from the cross section.
As is well-known, the cross section of the hadron pair production
being semi-inclusive process, can be calculated based on the parton model
with various fragmentation functions.  Then, by using symmetry relations
among fragmentation functions and taking an appropriate combination
of various hadron pair production processes, we can possibly remove the
contribution of QCD Compton components from the cross section
and thus, get clear information
of the polarized gluon distribution from the remaining PGF components.
Here, to show how to do this practically, we consider the light hadron pair
production with large transverse
momentum as shown in Fig. 1.


\section{Cross sections and double spin asymmetry
for large--$p_T$ pion pair production}

Let us consider the process of  $\ell + N \to \ell' + h_1 + h_2 +X$
in polarized lepton scattering off polarized nucleon targets
(Fig. 1), where $h_1$ and $h_2$ denote light hadrons in a pair.
As mentioned above,
the spin--dependent differential cross section at the leading order of
QCD can be given by the sum of the
PGF process and QCD Compton as follows,
\begin{equation}
d\Delta\sigma = d\Delta\sigma_{PGF} + d\Delta\sigma_{QCD}~.
\label{eqn:2}
\end{equation}
with
$$
d\Delta\sigma=d\sigma_{-+}-d\sigma_{++}+d\sigma_{+-}-d\sigma_{++}~.
$$
Here, for example, $d\sigma_{-+}$ denote that the helicity of
an initial lepton and the one of a target proton is
negative and positive, respectively.
Each term on the right hand side of eq.(\ref{eqn:2}) is given by
\begin{eqnarray}
d\Delta\sigma_{PGF}& \sim & \Delta g(\eta, Q^2)
d\Delta\hat\sigma_{PGF} \sum_{i=u, d, s, \bar u, \bar d, \bar s}
e_i^2 \{ D^{h_1}_i(z_1, Q^2) D^{h_2}_{\bar i}(z_2, Q^2) +
(1\leftrightarrow 2)\}~,\\
\label{eqn:3}
d\Delta\sigma_{QCD}& \sim & \sum_{q=u, d, s, \bar u, \bar d, \bar s}
e_q^2 \Delta f_q(\eta, Q^2) d\Delta\hat\sigma_{QCD}
\{ D^{h_1}_q(z_1, Q^2) D^{h_2}_g(z_2, Q^2) +
(1\leftrightarrow 2 )\}~,
\label{eqn:4}
\end{eqnarray}
where $\Delta g(\eta, Q^2)$, $\Delta f_q(\eta, Q^2)$ and
$D^h_i(z, Q^2)$ denote the polarized gluon and $q$--th quark distribution
functions with momentum fraction $\eta$ and the fragmentation
function of a hadron $h$ with momentum fraction $z$ emitted
from a parton $i$, respectively.
$d\Delta\hat\sigma_{PGF}$ and $d\Delta\hat\sigma_{QCD}$ are the
polarized differential cross sections of hard scattering subprocesses for
$\ell g\to \ell' q\bar q$ and 
$\ell \stackrel{\scriptscriptstyle(-)}{q}\to \ell' \stackrel{}{g}
\stackrel{\scriptscriptstyle(-)}{q}$ at the leading order QCD,
respectively.

Here we consider the following 4 pairs of a combination for the produced
hadrons $h_1$ with $z_1$ and $h_2$ with $z_2$,
$$
\rm{(i)}~~~~( \pi^+, \pi^- )~,~~~~~
\rm{(ii)}~~~~( \pi^-, \pi^+ )~,~~~~~
\rm{(iii)}~~~( \pi^+, \pi^+ )~,~~~~~
\rm{(iv)}~~~~( \pi^-, \pi^- )~,
$$
where (particle 1, particle 2) corresponds to ($h_1$ with $z_1$,
$h_2$ with $z_2$).
Then, the differential cross section of eq.(\ref{eqn:2}) for each pair
can be written as
\begin{eqnarray}
\rm{(i)}~~~~~~~~~~~& & \nonumber\\
d\Delta\sigma^{\pi^+\pi^-} &\sim& \Delta g(\eta, Q^2)
d\Delta\hat\sigma_{PGF} \left\{ \right.
\frac{4}{9}D^{\pi^+}_u(z_1, Q^2) D^{\pi^-}_{\bar u}(z_2, Q^2)+
\frac{1}{9}D^{\pi^+}_d(z_1, Q^2) D^{\pi^-}_{\bar d}(z_2, Q^2)\nonumber\\
& + &\frac{1}{9}D^{\pi^+}_s(z_1, Q^2) D^{\pi^-}_{\bar s}(z_2, Q^2)+
( \pi^+(z_1)\leftrightarrow\pi^-(z_2) )\left. \right\} \nonumber\\
& + & \frac{4}{9} \Delta u(\eta, Q^2) d\Delta\hat\sigma_{QCD}
\left\{ D^{\pi^+}_u(z_1, Q^2) D^{\pi^-}_g(z_2, Q^2) +
D^{\pi^-}_u(z_2, Q^2) D^{\pi^+}_g(z_1, Q^2)\right\} \nonumber\\
& + &({\rm contributions~~from~~}\Delta d, \Delta s,
\Delta\bar u, \Delta\bar d~~{\rm and}~~\Delta\bar s )~,
\label{eqn:5} \\
\rm{(ii)}~~~~~~~~~~~& & \nonumber\\
d\Delta\sigma^{\pi^-\pi^+} &\sim& \Delta g(\eta, Q^2)
d\Delta\hat\sigma_{PGF} \left\{ \right.
\frac{4}{9}D^{\pi^-}_u(z_1, Q^2) D^{\pi^+}_{\bar u}(z_2, Q^2)+
\frac{1}{9}D^{\pi^-}_d(z_1, Q^2) D^{\pi^+}_{\bar d}(z_2, Q^2)\nonumber\\
& + &\frac{1}{9}D^{\pi^-}_s(z_1, Q^2) D^{\pi^+}_{\bar s}(z_2, Q^2)+
( \pi^-(z_1)\leftrightarrow\pi^+(z_2) )\left. \right\} \nonumber\\
& + & \frac{4}{9} \Delta u(\eta, Q^2) d\Delta\hat\sigma_{QCD}
\left\{ D^{\pi^-}_u(z_1, Q^2) D^{\pi^+}_g(z_2, Q^2) +
D^{\pi^+}_u(z_2, Q^2) D^{\pi^-}_g(z_1, Q^2)\right\} \nonumber\\
& + &({\rm contributions~~from~~}\Delta d, \Delta s,
\Delta\bar u, \Delta\bar d~~{\rm and}~~\Delta\bar s )~,
\label{eqn:6} \\
\rm{(iii)}~~~~~~~~~~~& & \nonumber\\
d\Delta\sigma^{\pi^+\pi^+} &\sim& \Delta g(\eta, Q^2)
d\Delta\hat\sigma_{PGF} \left\{ \right.
\frac{4}{9}D^{\pi^+}_u(z_1, Q^2) D^{\pi^+}_{\bar u}(z_2, Q^2)+
\frac{1}{9}D^{\pi^+}_d(z_1, Q^2) D^{\pi^+}_{\bar d}(z_2, Q^2)\nonumber\\
& + &\frac{1}{9}D^{\pi^+}_s(z_1, Q^2) D^{\pi^+}_{\bar s}(z_2, Q^2)+
( \pi^+(z_1)\leftrightarrow\pi^+(z_2) )\left. \right\} \nonumber\\
& + & \frac{4}{9} \Delta u(\eta, Q^2) d\Delta\hat\sigma_{QCD}
\left\{ D^{\pi^+}_u(z_1, Q^2) D^{\pi^+}_g(z_2, Q^2) +
D^{\pi^+}_u(z_2, Q^2) D^{\pi^+}_g(z_1, Q^2)\right\} \nonumber\\
& + &({\rm contributions~~from~~}\Delta d, \Delta s,
\Delta\bar u, \Delta\bar d~~{\rm and}~~\Delta\bar s )~,
\label{eqn:7} \\
\rm{(iv)}~~~~~~~~~~~& & \nonumber\\
d\Delta\sigma^{\pi^-\pi^-} &\sim& \Delta g(\eta, Q^2)
d\Delta\hat\sigma_{PGF} \left\{ \right.
\frac{4}{9}D^{\pi^-}_u(z_1, Q^2) D^{\pi^-}_{\bar u}(z_2, Q^2)+
\frac{1}{9}D^{\pi^-}_d(z_1, Q^2) D^{\pi^-}_{\bar d}(z_2, Q^2)\nonumber\\
& + &\frac{1}{9}D^{\pi^-}_s(z_1, Q^2) D^{\pi^-}_{\bar s}(z_2, Q^2)+
( \pi^-(z_1)\leftrightarrow\pi^-(z_2) )\left. \right\} \nonumber\\
& + & \frac{4}{9} \Delta u(\eta, Q^2) d\Delta\hat\sigma_{QCD}
\left\{ D^{\pi^-}_u(z_1, Q^2) D^{\pi^-}_g(z_2, Q^2) +
D^{\pi^-}_u(z_2, Q^2) D^{\pi^-}_g(z_1, Q^2)\right\} \nonumber\\
& + &({\rm contributions~~from~~}\Delta d, \Delta s,
\Delta\bar u, \Delta\bar d~~{\rm and}~~\Delta\bar s )~.
\label{eqn:8}
\end{eqnarray}
Based on the isospin symmetry and charge conjugation invariance of
the fragmentation functions, various fragmentation functions in
eqs.(\ref{eqn:5})--(\ref{eqn:8})
can be classified into the following 4 functions\cite{Kumano},
\begin{eqnarray}
&&D\equiv D^{\pi^+}_u=D^{\pi^+}_{\bar d}=D^{\pi^-}_d=D^{\pi^-}_{\bar u}~~,
\nonumber\\
&&\widetilde{D}\equiv D^{\pi^+}_d=D^{\pi^+}_{\bar u}=D^{\pi^-}_u
=D^{\pi^-}_{\bar d}~~,\nonumber\\
&&D_s\equiv D^{\pi^+}_s=D^{\pi^+}_{\bar s}
=D^{\pi^-}_s=D^{\pi^-}_{\bar s}~~,\nonumber\\
&&D_g\equiv D^{\pi^+}_g=D^{\pi^-}_g~~,\nonumber
\end{eqnarray}
where $D$ and $\widetilde{D}$ are called favored and unfavored
fragmentation functions, respectively.
Considering the suppression of the $s$ quark contribution to
the pion production compared with the $u$ and $d$ quark
contribution, we do not identfy $D_s$ with $\widetilde{D}$.
This seems to be confirmed by 'leading paritcle'
measurements\cite{Kretzer,Abe}.
By using these 4 kinds of pion fragmentation functions,
we can make an interesting combination of cross sections which
contains only the PGF contribution as follows;
\begin{eqnarray}
& &d\Delta\sigma^{\pi^+\pi^-}+
d\Delta\sigma^{\pi^-\pi^+}-d\Delta\sigma^{\pi^+\pi^+}-
d\Delta\sigma^{\pi^-\pi^-} \sim \frac{10}{9}
\Delta g(\eta, Q^2)
d\Delta\hat\sigma_{PGF}
\label{eqn:13}\\
& & \times \left\{
D(z_1, Q^2) D(z_2, Q^2)+\widetilde{D}(z_1, Q^2) \widetilde{D}(z_2, Q^2)-
D(z_1, Q^2) \widetilde{D}(z_2, Q^2)-\widetilde{D}(z_1, Q^2) D(z_2, Q^2)
\right\}~.\nonumber
\end{eqnarray}
From this combination, we can calculate the double spin
asymmetry $A_{LL}$ defined by
\begin{eqnarray}
A_{LL}&=&
\frac{d\Delta\sigma^{\pi^+\pi^-}+
d\Delta\sigma^{\pi^-\pi^+}-d\Delta\sigma^{\pi^+\pi^+}-
d\Delta\sigma^{\pi^-\pi^-}}
{d\sigma^{\pi^+\pi^-}+
d\sigma^{\pi^-\pi^+}-d\sigma^{\pi^+\pi^+}-
d\sigma^{\pi^-\pi^-}}\nonumber\\
&=&\frac{\Delta g(\eta, Q^2)}{g(\eta,
Q^2)}\cdot\frac{d\Delta\hat\sigma_{PGF}}
{d\hat\sigma_{PGF}}~,
\label{eqn:14}
\end{eqnarray}
where the factor of the fragmentation function in eq.(\ref{eqn:13})
is dropped out from the numerator and the denominator of $A_{LL}$.
Therefore, from the measured $A_{LL}$, one can get clear
information of $\Delta g/g$ with reliable calculation of
$d\Delta\hat\sigma_{PGF}/d\hat\sigma_{PGF}$.
\footnote{Here we take account of the light quarks alone in
eqs.(\ref{eqn:5})--(\ref{eqn:8}).
For large $Q^2$ regions, heavy quarks might be generated 
for PGF process and QCD compton process might also have contributions
of heavy quarks.  Even then, the $A_{LL}$ is reduced to
eq.(\ref{eqn:14}) if
$D^{\pi^+}_Q=D^{\pi^+}_{\bar Q}
=D^{\pi^-}_Q=D^{\pi^-}_{\bar Q}$.}

Furthermore, when $z_1=z_2$, we have another formula
\begin{eqnarray}
& &d\Delta\sigma^{\pi^\pm\pi^\mp} -d\Delta\sigma^{\pi^+\pi^+}-
d\Delta\sigma^{\pi^-\pi^-} \sim \frac{10}{9}
\Delta g(\eta, Q^2)
d\Delta\hat\sigma_{PGF}
\label{eqn:15}\\
& & \times \left\{
D(z, Q^2) D(z, Q^2)+\widetilde{D}(z, Q^2) \widetilde{D}(z, Q^2)-
2~D(z, Q^2) \widetilde{D}(z, Q^2)\right\}~.\nonumber
\end{eqnarray}
In this case, we can also define the double spin
asymmetry $A_{LL}$ as follows;
\begin{equation}
A_{LL}=
\frac{d\Delta\sigma^{\pi^{\pm}\pi^{\mp}}
-d\Delta\sigma^{\pi^+\pi^+}-d\Delta\sigma^{\pi^-\pi^-}}
{d\sigma^{\pi^{\pm}\pi^{\mp}}
-d\sigma^{\pi^+\pi^+}-
d\sigma^{\pi^-\pi^-}}
=\frac{\Delta g(\eta, Q^2)}{g(\eta, Q^2)}\cdot\frac{d\Delta\hat\sigma_{PGF}}
{d\hat\sigma_{PGF}}~.
\label{eqn:16}
\end{equation}
The $A_{LL}$ defined in eq.(\ref{eqn:14}) and eq.(\ref{eqn:16})
results in the same physical quantity.  From eq.(\ref{eqn:16}),
we can also extract $\Delta g/g$ as well as in eq.(\ref{eqn:14}).

\section{Numerical calculations of the cross sections and the double
spin asymmetry}

In this section, to see how the above formula works, we numerically
calculate the double spin asymmetry
for the large--$p_T$ pion pair production of pol--DIS.
The spin--independent (spin--dependent) differential cross sections
for producing hadrons $h_1$ and $h_2$ are given by\cite{Peccei}
\begin{equation}
\frac{d( \Delta )\sigma^{h_1 h_2}}
{dz_1d\cos\theta_1dz_2d\cos\theta_2 dx dy}=
\frac{d( \Delta )\sigma^{h_1 h_2}_{PGF}}
{dz_1d\cos\theta_1dz_2d\cos\theta_2 dx dy} +
\frac{d( \Delta )\sigma^{h_1 h_2}_{QCD}}
{dz_1d\cos\theta_1dz_2d\cos\theta_2 dx dy}~.
\label{eqn:17}
\end{equation}
Each term in the right hand side of eq.(\ref{eqn:17}) is written as
\begin{eqnarray}
\frac{d( \Delta )\sigma^{h_1 h_2}_{PGF}}
{dz_1d\cos\theta_1dz_2d\cos\theta_2 dx dy} &=&
(\Delta )g(\eta, Q^2)~C(\theta_1, \theta_2)
\frac{d( \Delta )\widehat{\sigma}^{h_1 h_2}_{PGF}}
{dz_id\cos\theta_1dz_{\bar i}d\cos\theta_2 dx_g dy}\nonumber\\
&\times& \sum_{i=u, d, s, \bar u, \bar d, \bar s}
e_i^2 \{ D^{h_1}_i(z'_1, Q^2) D^{h_2}_{\bar i}(z'_2, Q^2) +
(1\leftrightarrow 2)\}~,\\
\label{eqn:18}
\frac{d( \Delta )\sigma^{h_1 h_2}_{QCD}}
{dz_1d\cos\theta_1dz_2d\cos\theta_2 dx dy} &=&
\sum_{q=u, d, s, \bar u, \bar d, \bar s} e_q^2
(\Delta )f_q(\eta, Q^2)~C(\theta_1, \theta_2)
\frac{d( \Delta )\widehat{\sigma}^{h_1 h_2}_{QCD}}
{dz_id\cos\theta_1dz_gd\cos\theta_2 dx_q dy}\nonumber\\
&\times& \{ D^{h_1}_q(z'_1, Q^2) D^{h_2}_g(z'_2, Q^2) +
(1\leftrightarrow 2)\}~,
\label{eqn:19}
\end{eqnarray}
where
\begin{eqnarray}
&& \eta=x+(1-x)\tau_1\tau_2~,~~~~~Q^2 = x~y~s~,\nonumber\\
&&z'_1=\left (\frac{\tau_1+\tau_2}
{\tau_2}\right )z_1~,~~~~~
z'_2=\left (\frac{\tau_1+\tau_2}
{\tau_1}\right )z_2~,
\label{eqn:20}\\
&&C(\theta_1, \theta_2) =
\frac{(\tau_1+\tau_2)^2}
{\eta~\tau_1\tau_2}~
\frac{(1-x)}{8\cos ^2\frac{1}{2}\theta_1\cos ^2\frac{1}{2}\theta_2
\sin ^2\frac{1}{2}(\theta_1+\theta_2)}~,\nonumber
\end{eqnarray}
with
$$
\tau_{1, 2}=\tan\frac{1}{2}\theta_{1, 2}~.
$$
Here we simply assume the scattering angle of outgoing hadrons
$\theta_{1,~2}$ to be the same with the one of scattered partons
in the virtual photon--nucleon c.m. frame.
This assumption might not be unreasonable if observed particles
are light hadrons with high energy.
$s$ is the total squared energy of the lepton scattering off the nucleon.
$x$, $y$ and $z_{1,~2}$ in eqs.(\ref{eqn:17})--(\ref{eqn:20}) are
familiar kinematic variables for semi--inclusive processes in DIS and
are defined as
$$
x=\frac{Q^2}{2P\cdot q}~,~~~~~y=\frac{P\cdot q}{P\cdot\ell}~,~~~~~
z_{1,~2}=\frac{P\cdot P_{1,~2}}{P\cdot q}~,
$$
where $\ell$, $q$, $P$ and $P_{1,2}$ are the momentum of the incident
lepton,
virtual photon, target nucleon and outgoing hadrons, respectively.
The differential cross sections of hard scattering subprocesses
with outgoing two partons having opposite in an azimuth angle for
$\ell g\to \ell' q\bar q$ and 
$\ell \stackrel{\scriptscriptstyle(-)}{q}\to \ell' \stackrel{}{g}
\stackrel{\scriptscriptstyle(-)}{q}$ at the leading order QCD
are given as
\begin{eqnarray}
\frac{d(\Delta)\hat\sigma_{PGF (QCD)}}
{dz_id\cos\theta_1d\phi_1dz_{\bar i (g)}d\cos\theta_2 dx_{g (q)}
dy d\phi_{\ell}}&=&
\frac{1}{128\pi^2}\frac{\alpha^2\alpha_s}{(p_0\cdot\ell)Q^2}
\frac{y(\eta-x)(1-\eta)^2}{x}~\nonumber\\
&\times&~B(\theta_1, \theta_2)~e_{\ell}^2~e_i^2|(\Delta)M|^2_{PGF (QCD)}~,
\label{eqn:sub}
\end{eqnarray}
with
$$
\frac{1}{B(\theta_1, \theta_2)}=
\sin (\theta_1+\theta_2)
\left [\frac{\{z_i(1-\eta)+(\eta-x)\}\sin\theta_1+
\{z_{\bar i(g)}(1-\eta)+(\eta-x)\}\sin\theta_2}
{\sin\theta_1\sin\theta_2}\right ]~,
$$
where $z_i$, $z_{\bar i}$ and $z_g$ are the momentum fraction of
the outgoing parton $i$, $\bar i$ and $g$, respectively, to the incoming
parton, and are given as\cite{Peccei}
$$
z_i=\frac{\tau_2}{\tau_1+\tau_2}~,~~~
z_{\bar i (g)}=\frac{\tau_1}{\tau_1+\tau_2}~.
$$
The amplitude $|(\Delta)M|^2_{PGF (QCD)}$ in eq.(\ref{eqn:sub}) is
given by \cite{Mirkes}
\begin{eqnarray}
|M|^2_{PGF}&=&16(\ell\cdot\ell')\left [
\frac{(\ell\cdot p_1)^2+(\ell'\cdot p_1)^2+
(\ell\cdot p_2)^2+(\ell'\cdot p_2)^2}{(p_0\cdot p_1)(p_0\cdot
p_2)}\right ]~,
\nonumber\\
|M|^2_{QCD}&=&\frac{128}{3}(\ell\cdot\ell')\left [
\frac{(\ell\cdot p_0)^2+(\ell'\cdot p_0)^2+
(\ell\cdot p_1)^2+(\ell'\cdot p_1)^2}{(p_0\cdot p_2)(p_1\cdot p_2)}\right ]~
\nonumber
\end{eqnarray}
for the spin--independent case and
\begin{eqnarray}
|\Delta M|^2_{PGF}&=&16(\ell\cdot\ell')\left [
\frac{(\ell'\cdot p_1)^2-(\ell\cdot p_1)^2+
(\ell'\cdot p_2)^2-(\ell\cdot p_2)^2}{(p_0\cdot p_1)(p_0\cdot
p_2)}\right ]~,
\nonumber\\
|\Delta M|^2_{QCD}&=&
\frac{128}{3}(\ell\cdot\ell')\left [
\frac{(\ell\cdot p_0)^2-(\ell'\cdot p_0)^2-
(\ell\cdot p_1)^2+(\ell'\cdot p_1)^2}{(p_0\cdot p_2)(p_1\cdot p_2)}\right ]~
\nonumber
\end{eqnarray}
for the spin--dependent case.
The $p_0$, $p_1$, $p_2$ and $\ell'$ are the momentun of the incoming parton,
outgoing parton $i$, $\bar i$, $g$ and the outgoing lepton, respectively,
and are written by
\begin{eqnarray}
&&p_0^{\mu}=|P|(\eta~,~0~,~0~,~-\eta)~,~~~
p_1^{\mu}=|p_1|(1~,~\sin\theta_1\cos\phi_1~,~
\sin\theta_1\sin\phi_1~,~\cos\theta_1)~,\nonumber\\
&&p_2^{\mu}=|p_2|(1~,~\sin\theta_2\cos(\phi_1-\pi)~,~
\sin\theta_2\sin(\phi_1-\pi)~,~\cos\theta_2)~,\nonumber\\
&&\ell^{\mu}=|\ell|(1~,~\sin\theta_{\ell}\cos\phi_{\ell}~,~
\sin\theta_{\ell}\sin\phi_{\ell}~,~\cos\theta_{\ell})~,\nonumber\\
&&q^{\mu}=|P|(1-2x~,~0~,~0~,~1)~,~~~
\ell'_{\mu}=\ell_{\mu}-q_{\mu}~,\nonumber
\end{eqnarray}
with
\begin{eqnarray}
&&|P|=\sqrt{\frac{Q^2}{4x(1-x)}}~,~~~
|p_{1, 2}|=|P|\{z_{i, \bar i(g)}(1-\eta)+(\eta-x)\}~,\nonumber\\
&&\cos\theta_{1, 2}=\frac{z_{i, \bar i(g)}(1+\eta-2x)-(\eta-x)}
{z_{i, \bar i(g)}(1-\eta)+(\eta-x)}~,~~~
|\ell|=|P|\frac{1-xy}{y}~,~~~
\cos\theta_{\ell}=\frac{1-2x+xy}{1-xy}~,\nonumber
\end{eqnarray}
in the c.m. frame of the virtual photon--nucleon system\cite{Peccei}.
By using these formulas and newly analyzed pion fragmentation
functions\cite{Kretzer},
we have calculated the spin--dependent and
spin--independent cross sections of the large--$p_T$ pion pair
prodcution and estimated the double spin asymmetry of
eq.(\ref{eqn:14}) at the energy of HERMES experiments.
Here, we have taken the AAC\cite{Goto} and GS96\cite{GS96}
parametrizations at LO QCD as polarized parton distribution functions
and GRV98\cite{GRV98} and MRST98\cite{MRST98} as unpolarized ones.
At $\sqrt s=7.25$GeV, $y=0.75$, $Q^2\geq 1$GeV$^2$ and $W^2\geq 10$GeV$^2$
with two different sets of kinematical
values of $\theta_{1,2}$ and $z_{1,2}$ for
the produced pion pair, the calculated results of the spin--independent
(spin--dependent) differential cross sections and $A_{LL}$ are shown
as a function of $\eta$ in Figs. 2 and 3, respectively.
\footnote{Here the polarized parton distributions were evolved from
$Q_0^2=1$GeV$^2$ to any $Q^2$ value.  Though the initial 
$Q^2_0$ value of the GS96 model is taken to be 4GeV$^2$ in original 
literature, we simply assumed their parton distributions to 
be in scaling for $1\leq Q^2<4$GeV$^2$.}
From Fig. 3, one can see a big difference of the behavior
of $A_{LL}$ depending on the models of  $\Delta g/g$ and hence,
we can extract the behavior of $\Delta g$
rather clearly from this analysis.


\section{Conclusion and discussion}

We proposed a new formula for extracting the polarized gluon distribution
from the large--$p_T$ light hadron pair production
in pol--DIS by making an appropriate combination
of hadron pair productions.
Since the double spin asymmetry $A_{LL}$ for this combination
is directly proportinal to
$\Delta g/g$, the measurement of this quantity is quite promising for
getting rather clear information on the polarized gluon
distribution in the nucleon.

In this work, we calculated only the case of
the large--$p_T$ pion pair production.
The same analysis can be applied also for the kaon or the proton pair
productions by considering the reflection symmetry along the
V--spin axis, the isospin symmetry and charge conjugation invariance of
the fragmentation functions as follows\cite{Kumano}
\begin{eqnarray}
&&D^K_s\equiv D^{K^+}_{\bar s}=D^{K^-}_s~~,\nonumber\\
&&\widetilde{D^K_s}\equiv D^{K^+}_s=D^{K^-}_{\bar s}~~,\nonumber\\
&&D^K_u\equiv D^{K^+}_u=D^{K^-}_{\bar u}~~,\nonumber\\
&&\widetilde{D^K_{u,d}}\equiv D^{K^+}_{\bar u}=D^{K^+}_d =
D^{K^+}_{\bar d}=D^{K^-}_u=D^{K^-}_d =D^{K^-}_{\bar d}~~,\nonumber\\
&&D^K_g\equiv D^{K^+}_g=D^{K^-}_g\nonumber
\end{eqnarray}
for the kaon pair production case and
\begin{eqnarray}
&&D^p\equiv D^p_u=D^p_d=D^{\bar p}_{\bar u}=D^{\bar p}_{\bar
d}~~,\nonumber\\
&&\widetilde{D^p}\equiv D^p_{\bar u}=D^p_{\bar d}=
D^{\bar p}_u=D^{\bar p}_d~~,\nonumber\\
&&D^p_s\equiv D^p_s=D^p_{\bar s}=D^{\bar p}_s=D^{\bar p}_{\bar
s}~~,\nonumber\\
&&D^p_g\equiv D^p_g=D^{\bar p}_g\nonumber
\end{eqnarray}
for the proton pair production case.

In general, it might not be so easy to precisely extract a physical quantity
from a combination of data on the different kind of physical quantities.
However, if we have enough data with high precision for
wide kinematical region, it could be possible to determine
the physical quantity concerned with rather precisely from
those data, just as in the case of Bjorken sum rule in which
two different quantities, $g_1^p$ and $g_1^n$, were combined.  We believe
that the formulas proposed here are suitable for those analyses,
since we can expect to have rather many and precise data
because of large cross sections for the light hadron pair production.
The HERMES RICH detector at DESY which will provide good particle
identification in the wide momentum range
could allow to measure $A_{LL}$ for the processes discussed here.

\vspace{1em}
This work is supported by the Grant--in--Aid for Science Research,
Ministry of Education, Science and Culture, Japan (No.11694081).

\newpage


\begin{figure} 
\begin{center}
\epsfig{figure=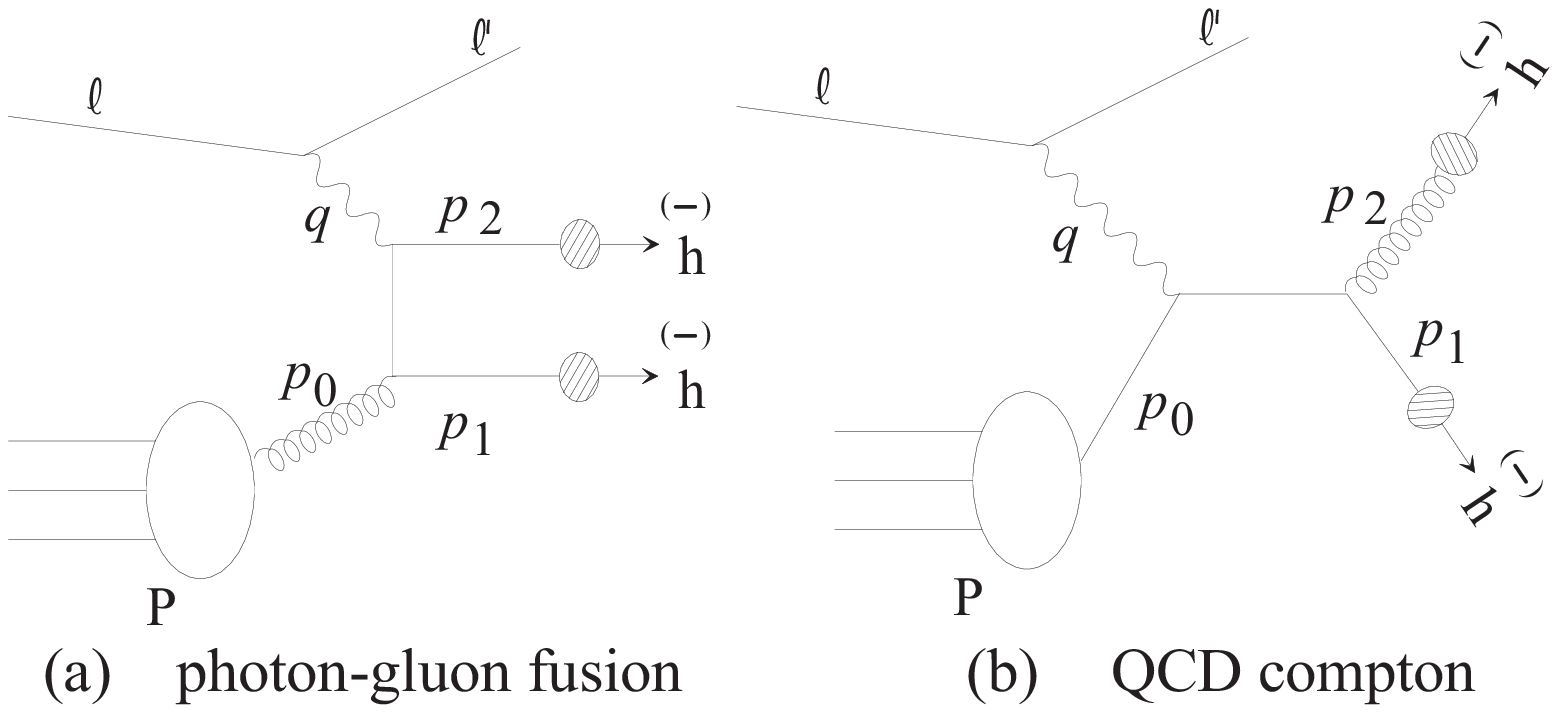,height=6.5cm}
\caption{Lowest order Feynman diagrams for the large--$p_T$
         hadron pair production.}
\end{center}
\end{figure}

\vspace{2em}

\begin{figure} 
\centerline{\epsfig{file=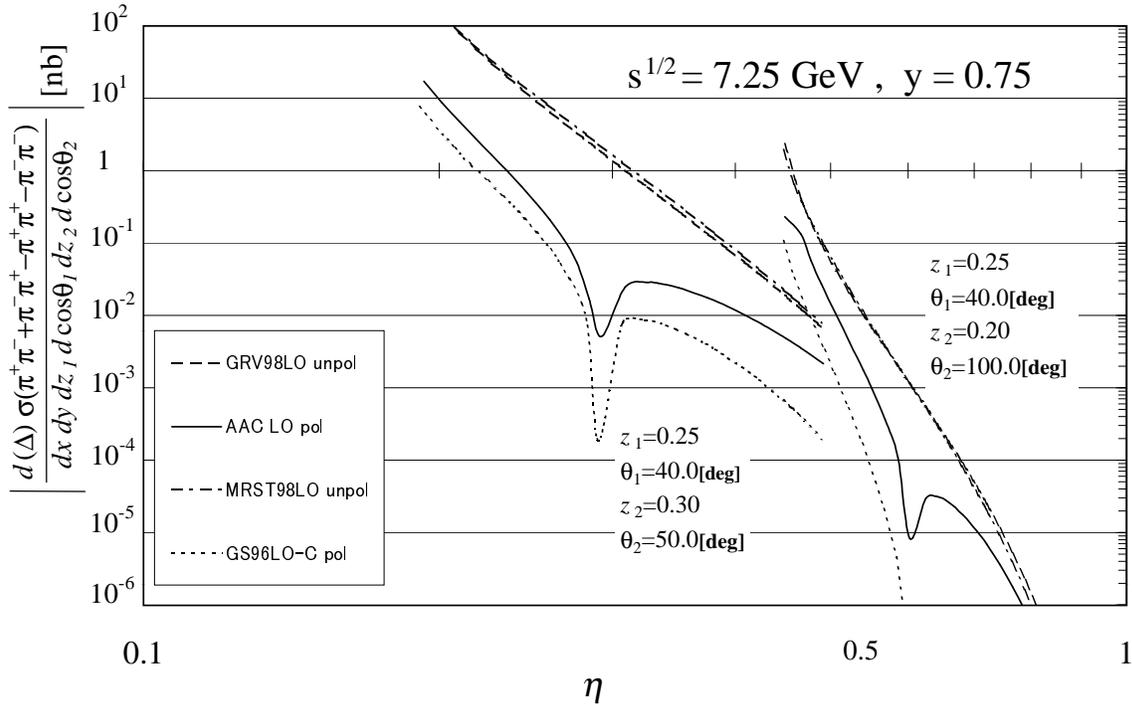,height=13cm}}
\vspace{10pt}
\caption{Combined spin--independent and spin--dependent
differential cross sections defined at the denominator and numerator,
respectively, of eq.(10) as a function of $\eta$ at 
$\sqrt{s} = 7.25$ GeV, $y=0.75$ for the deep inelastic 
regions ($Q^2\geq 1$GeV$^2$ and $W^2\geq 10$GeV$^2$) with 
two different sets of kinematical values($\theta_{1,2}$, 
$z_{1,2}$) of the produced pion pair.
$\eta$ is the momentum fraction of the gluon and
obtained from $x$ by eq.(16).}
\label{fig2}
\end{figure}

\newpage
\begin{figure}[t!] 
\centerline{\epsfig{file=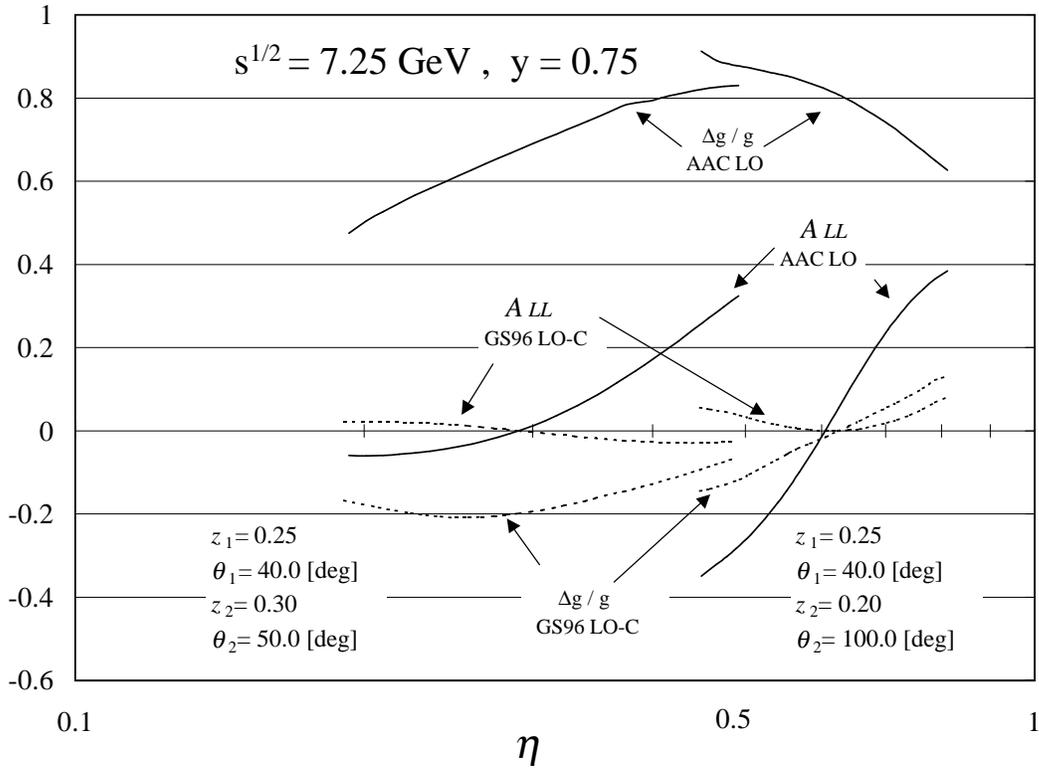,height=13cm}}
\vspace{10pt}
\caption{The $\eta$ dependence of $A_{LL}$ at $\sqrt{s} = 7.25$
GeV, $y=0.75$ for two different sets of kinematical values($\theta_{1,2}$, 
$z_{1,2}$) of the produced pion pair.  Solid line and dotted line are 
for AAC LO and GS96LO--C parametrization models, respectively.
$\Delta g/g$ itself is also presented for both parametrization models.} 
\label{fig3}
\end{figure}

\end{document}